\documentclass[prd, nobibnotes, nofootinbib,
onecolumn,10pt, a4paper, superscriptaddress, nofootinbib,
byrevtex]{revtex4}

\usepackage{psfig}
\usepackage{graphics}

\def\beq{\begin{equation}}
\def\eeq{\end{equation}}
\def\bea{\begin{eqnarray}}
\def\eea{\end{eqnarray}}
\newcommand{\dle}[1]{\label{#1}}

\newcommand{\dr}[1]{\ref{#1}}

\def\tE{\tilde{E}}
\def\hl{\hat{\ell}}
\def\bq{\begin{quote}}
\def\eq{\end{quote}}
\def\bi{\begin{itemize}}
\def\ei{\end{itemize}}
\def\beqa{\begin{eqnarray}}
\def\eeqa{\end{eqnarray}}
\def\be{\begin{enumerate}}
\def\ee{\end{enumerate}}
\def\beq{\begin{equation}}
\def\eeq{\end{equation}}
\def\bi{\begin{itemize}}
\def\ei{\end{itemize}}

\def\r2{\sqrt{2}}

\def\bi{\begin{itemize}}
\def\ei{\end{itemize}}

\def\nn{\nonumber \\}
\newcommand{\gsim}{\raise.3ex\hbox{$>$\kern-.75em\lower1ex\hbox{$\sim$}}}
\newcommand{\lsim}{\raise.3ex\hbox{$<$\kern-.75em\lower1ex\hbox{$\sim$}}}

\begin{document}

\title{A comment on bouncing and cyclic branes in more than one
extra-dimension}

\author{Ph.Brax}
\email{brax@spht.saclay.cea.fr} \affiliation{Service de Physique
Th\'eorique, CEA Saclay, F--91191 Gif-sur-Yvette, France}

\author{D.A.Steer}
\email{steer@th.u-psud.fr} \affiliation{Laboratoire de Physique
Th\'eorique, B\^at. 210, Universit\'e Paris XI, F--91405 Orsay
Cedex, France} \affiliation{F\'ed\'eration de Recherche APC,
Universit\'e Paris VII, France.}

\vspace*{0.5cm}

\begin{abstract}

We argue that bouncing branes occur naturally when there is more
than one extra-dimension. We consider three-branes embedded in
space-times with a horizon and an isometry group $SO(6)$. As soon
as the brane angular momentum is large enough, a repulsive barrier
prevents the branes from reaching the horizon. We illustrate this
phenomenon with the case of D3-branes in an
AdS$_5$-Schwarzschild$\times S_5$ background and asymptotically
flat space-time.

\vspace*{0.5cm} \noindent {\footnotesize Preprint numbers:
SPhT-Saclay T02/094; ORSAY-LPT-02-74}
\end{abstract}


\maketitle

Recently there has been renewed interest in the study of
oscillating (cyclic) and bouncing universes \cite{Neil1, Neil2}
motivated in part by different brane world scenarios \cite{ovr}.
In these studies only one extra dimension is typically considered,
and matter on the brane is neglected so that the motion is solely
induced from the bulk. When there is one extra dimension (or
equivalently, when the brane is only free to move in one extra
dimension), a number of different bulk space-times have been shown
to lead to bouncing and oscillating universes, for instance in the
case of  a charged AdS$_5$ blackhole \cite{Peloso} and the
Klebanov-Strassler background \cite{Kackru}.

The purpose of this short note is simply to comment on the generic
occurrence of bouncing branes when there is more than one extra
dimension in which the brane can move. To illustrate this point we
consider the example of a brane moving in
AdS$_5$-Schwarzschild$\times$S$_5$. The dynamics are parametrised
by a conserved angular momentum on $S_5$ \cite{KK} reminiscent of
a point particle moving in a central well. If the angular momentum
is large enough a centrifugal potential develops at small scales
off which the brane universe `bounces'. Furthermore if the spatial
sections of the brane have positive curvature then, as we
illustrate below, a repulsive potential also develops at large
scales.  The brane may then be bound between the two `bumps'
leading to an oscillating or cyclic universe \cite{Neil3}. We also
consider the case of asymptotically flat universes where the same
bouncing behaviour is generic.

Consider a D3-brane moving in Sch-AdS$_5\times S_5$ spacetime
which is the near horizon limit of a BPS $D3$-black-brane.  The
bulk metric in the Einstein frame is given by
\bea
ds^2 & \equiv &  g_{00}(r) dt^2 + g_d(r)\chi_{ij} dx^i dx^j +
g_{rr}(r)dr^2 + g_s(r) d \Omega_5^2
\nn
&=& \frac{r^2}{L^2} \left(
-f(r) dt^2 + \chi_{ij} dx^i dx^j \right) + \frac{L^2 dr^2}{f(r)
r^2 } + L^2 d\Omega_5^2
\dle{AdS}
\eea
where $L$ is the AdS curvature, $\chi_{ij}$ ($i=1,2,3$)
is the metric on a space of constant curvature $k=1,0,-1$
(closed, flat, open respectively), and
\beq
f(r) = k\frac{L^2}{r^2} + 1- \left( \frac{r_H}{r} \right)^4.
\dle{fdef}
\eeq
The metric (\dr{AdS}) is a solution of the ten dimensional
supergravity  equations with corresponding bulk 4-form field
\beq
C_{0123} = \sqrt{\chi} \left( - \left(\frac{r}{L}\right)^4 +
\frac{r_H^4}{2L^4} \right)
\dle{Cdef}
\eeq
and a constant dilaton $\Phi$ (which we set to zero).  It will be
convenient to work with the dimensionless variables
\beq
y = \frac{r}{L} \qquad , \qquad r_0=\frac{r_H}{L}
\eeq
so that for $k=0$ the horizon is at $y=r_0$.  When $k =\pm 1$ the
horizon is at $2y = -k + (1+4r_0^4)^{1/2}$.  Later we will comment
briefly on the case $r_0 = 0, k=- 1$, namely open AdS$_5 \times
S_5$ which has a horizon at $y=1$.

The D3-brane dynamics are governed by the Dirac-Born-Infeld action
\begin{equation}
S =-T  \int d^{4} x e^{-\Phi} \sqrt { - \det (\gamma_{\mu \nu} ) }
- q T \int d^4 x \hat{C}_4 \dle{action}
\end{equation}
where $T$ is the brane tension, $q=(-)1$ for a BPS (anti-) brane,
and $\gamma_{\mu \nu}$,  $\mu=0\dots 3$ is the induced brane
metric in  the string frame. As mentioned above we set all matter
on the brane to zero so that the pull-back of the Neveu-Schwarz
anti-symmetric two-form $\hat{B}_{\mu \nu}$ as well as the
worldvolume anti-symmetric gauge fields $F_{\mu \nu}$ are assumed
to vanish. In the static gauge, the induced metric is given by
\beq
\gamma_{00} = e^{\Phi/2} \left( g_{00} + g_{rr}\dot{r}^2 + g_s
\dot{\phi}^2 \right) ,\qquad \gamma_{ij} = e^{\Phi/2} g_d
\chi_{ij}
\dle{static}
\eeq
where $\cdot = d/dt$, $\dot{\varphi}^2 = h_{pq}\dot{\varphi}^{p}
\dot{\varphi}^{q}$ with $h_{pq}$ ($p=1,\ldots,5$) the metric on
the 5-sphere, and $r(t), \phi(t)$ are the brane position at time
$t$. These can therefore be determined through the Lagrangian
${\cal L}$ defined by
\beq
S  = T  V_3 \int dt {\cal L} \; , \qquad  \qquad
 {\cal L}  =  - \sqrt{a+ b\dot{r}^2 + c
\dot{\varphi}^2 } + e
\dle{ssll}
\eeq
where $V_3 = \int d^3 x \sqrt{-\chi} $ is the spatial volume of
the probe, and
\beq
a\equiv -g_{00}g_d^3 = y^8 f(y) \; , \qquad b\equiv  -
g_{rr}g_d^3 = -\frac{y^4}{f}, \qquad c\equiv  - g_d^3 g_s = -y^6
L^2 \;, \qquad e\equiv -q \hat{C}_4 = q(y^4-\frac{r_0^4}{2}).
\dle{abcddef}
\eeq
(The backreaction of the brane on the bulk is neglected
\cite{KK}.) As discussed in \cite{KK}, it follows from
(\dr{ssll}) that there is a corresponding conserved positive
energy $E$ and angular momentum $\ell$ of the brane around the
$S^5$ from which
\beq
\dot{\varphi}^2  =
 \frac{a^2 \ell^2}{c^2(E+e)^2} \; , \qquad
\dot{r}^2  =  -\frac{a}{b} \left[ 1 + \frac{a}{c} \frac{ (\ell^2 -
c)}{(E+e)^2} \right].
\dle{Rdot}
\eeq
For an observer living on the brane, however, the brane time
$\tau$ and the scale factor $S(\tau)$ are given by
\beq
ds_{3}^2 =  - d\tau^2 + S^2(\tau) \chi_{ij} dx^i dx^j \qquad
\longleftrightarrow \qquad d\tau^2 = - \gamma_{00}(t) dt^2 \qquad
, \qquad S^2 = g_d(t) = y^2 = (r/L)^2
\dle{Sdef}
\eeq
From Eqs.~(\ref{Rdot}) and (\ref{Sdef}) we find $d\tau^2 = a^2
dt^2/(E+e)^2 g_d^3$ so that
\begin{equation}
\left(\frac{dr}{d\tau}\right)^2 = - \frac{1}{abc}g_d^3 \left[
c(E+e)^2 + a (\ell^2 - c) \right].
\dle{Rdash}
\end{equation}
Thus the Friedman equation on the brane, $H=
\frac{1}{S}\frac{dS}{d\tau}$, is given by
\begin{equation}
H^2=
 \frac{q^2-1}{L^2} - \frac{k}{a^2} + \frac{r_0^4 L^2}{a^4} +
 L^6   \left( \frac{\tilde{E}}{a^4} \right)
\left( \frac{\tilde{E}}{a^4} + \frac{2q}{L^4} \right) + \,
\ell^2\, \frac{ L^4}{a^6} \left(- \frac{k}{a^2}+  \frac{r_0^4 L^2}{a^4} -
\frac{1}{L^2} \right)
\dle{fred}
\end{equation}
where $a(\tau) = S(\tau) L = r(\tau)$ is the {\it dimensionful}
scale factor (not to be confused with the parameter $a$ in
(\dr{abcddef})) and $\tilde{E} = E - qr_0^4/2$. The
$\ell$-independent terms are familiar from non-$Z_2$-symmetric
brane world cosmology \cite{CU} with $\tilde{E}$ being a measure
of $Z_2$ symmetry breaking \cite{SP}. Note in (\dr{fred}) the new
$\ell$ dependent terms coming from the $S_5$ whose effect we want
to study. Observe also that the effective cosmological constant on
the brane vanishes for BPS branes $q=\pm 1$.  In the context of
brane-worlds, this is  the Randall-Sundrum fine tuning condition
for the bulk and brane cosmological constants. As in four
dimensional cosmology there is a curvature term. Moreover there is
also a ``dark radiation'' term arising from the existence of a
horizon in ten dimensions.  From now on we focus only on BPS
branes.

Rather than solving directly for $a(\tau)$ from (\dr{fred}) it is
simpler to consider the effective potential for the brane motion,
$V^{\tau} = E - \frac{1}{2} (dr/d\tau)^2$, which is given by
\beq
V^{\tau}_{\bf eff}(k,y,\hat{\ell},E)= E + \frac{1}{2y^6}
 \left[ f
y^2 \left( \hat{\ell}^2 + y^6 \right) - \left( \tilde{E} + q y^4
\right)^2 \right] \dle{Veffeta}
\eeq
where $f=f(k,y)$ is given in (\dr{fdef}) and we have defined the
dimensionless angular momentum
\beq
\hat{\ell} = \frac{\ell}{L}.
\eeq
This is just the classical problem of particle in a central
potential with force $F=-\partial V^{\tau}_{\bf eff}/\partial y$.
It follows from (\dr{Veffeta}) that for $q=\pm 1$
\begin{equation}
\lim_{r \rightarrow \infty}
 V^{\tau}_{\bf eff} = E + \frac{k}{2}.
\dle{limit}
\end{equation}
Thus if $k=0,-1$ the brane may reach infinity since $V^{\tau}_{\bf
eff} \leq E$ there. However, a closed universe $k=1$ will be
unable to go to infinity, {\it i.e.\ }there is an upper bound on
the brane scale factor $S=y$. At the horizon, on the other hand,
$f=0$ so that from (\dr{Veffeta}) $V^{\tau}_{\bf eff} < E$
independently of $k$ meaning that the vicinity of the horizon is
an allowed region. Finally, notice that the coefficient of the
$\hat{\ell}^2$ is positive and given by $ f y^2/2$.  Thus for
sufficiently large $\hat{\ell}^2$ and in some intermediate range
of $y$, we expect $V^{\tau}_{\bf eff} > E$ which is the
centrifugal barrier.  This behaviour is shown in figure 1a for
$k=0$ and in figure 1b for $k=1$.
\begin{figure}
\centerline{ \scalebox{0.8}{
\includegraphics{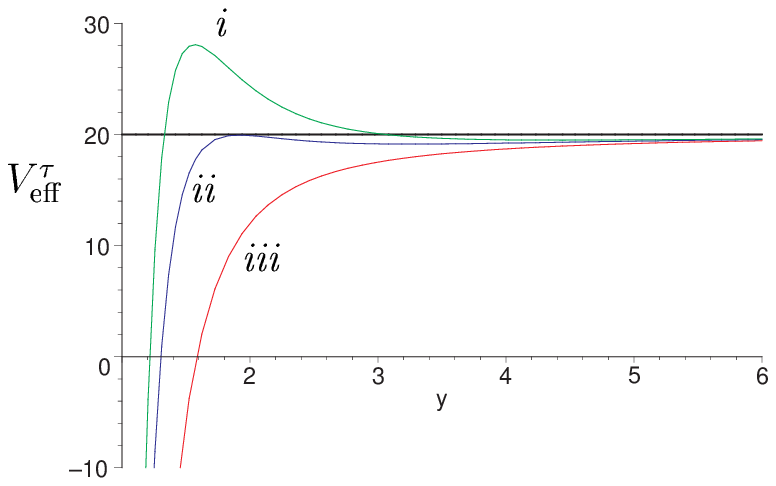}}
\hspace*{1cm} \scalebox{0.8}{ \includegraphics{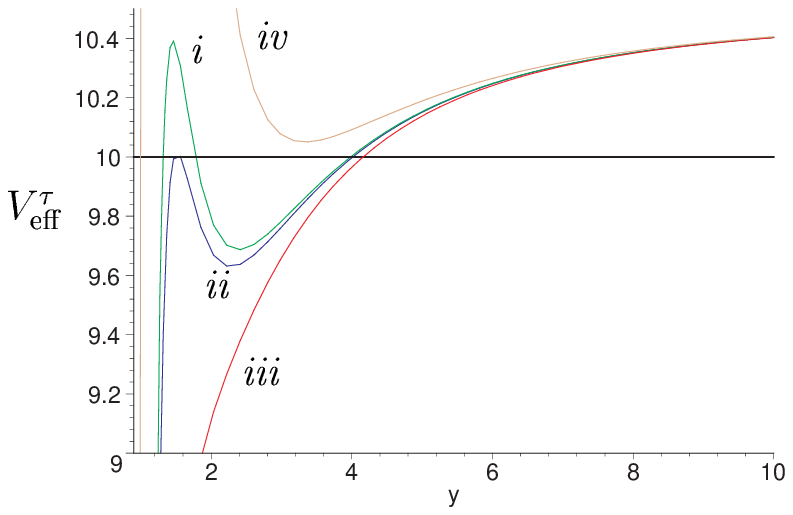}}}
\caption{LH panel: $V^{\tau}_{\rm eff}$ for a flat universe, $k=0$
and $E=20$.  RH panel: $V^{\tau}_{\rm eff}$ for a closed universe,
$k=1$ and $E=10$. In each case $q=+1$ and the central curve (blue,
labeled {\it ii}) has $\hl = \hl_c$ ($\sim 16$ for $k=0$ and
$\sim 8$ for $k=1$), the lower curve (red, labeled {\it iii}) has
$\hl < \hl_c$, and the upper curve (green, labeled {\it i}) has
$\hl
>\hl_c$.  Note that in the LH panel $V^{\tau}_{\rm eff}
\rightarrow E$ as $r\rightarrow \infty$ as expected from
(\dr{limit}).  In the RH panel $V^{\tau}_{\rm eff} \rightarrow E +
1/2$. Notice also the uppermost curve (brown, labeled {\it iv})
in the RH panel: if $\ell$ is sufficiently large then there is
neither a bounce nor a cyclic universe.}
\dle{fig1}
\end{figure}

Therefore for $k=0,-1$, an incoming (contracting) universe will
bounce off the centrifugal potential if $\ell > \ell_c$, where
$\ell_c$ is the critical angular momentum for which $V^{\tau}_{\rm
eff}=E$ (see figure \dr{fig1} and below). The bounce occurs at an
$\ell$-dependent minimum value of $S_m(\tau)=y_m(\tau)$ after
which the brane moves radially outwards and expands.  For
$\ell=\ell_c$ the brane reaches a circular orbit with
corresponding constant scale factor $y_c$, and for $\ell < \ell_c$
the brane falls into the horizon. For $k=1$, on the contrary, the
bounce turns into a cyclic or oscillating universe (see figure
\dr{fig1}b) since the brane may not reach infinity due to
(\dr{limit}).

In some cases the equations of motion are integrable.  Consider
the brane trajectory $r(\phi)$. Using equations (\dr{Rdot}), as
well as the definition of $a$, $b$, $c$ and $e$ in (\dr{abcddef})
leads to (for $q^2 = 1$)
\begin{equation}
\phi= \pm \frac{\hat{\ell}}{2}\int_{z(\phi)}^{z_0}\frac{dz}{[-kz^4 +
2qE z^3 - \hat{\ell}^2  z^2 + (\tE^2 - k \hat{\ell}^2) z+
\hat{\ell}^2r_0^4]^{1/2}}, \dle{parametric}
\end{equation}
where $z=y^2$, we have set $\phi(z_0) = 0$ where $z_0$ is the
initial position of the brane, and the $+/-$ sign corresponds to
an initially ingoing/outgoing brane. (These are initial
conditions.)
\\
\\
{\it Flat universe: $k=0$}
\\
When $k=0$ a bounce can occur.  Let $z_0 \rightarrow \infty$ and
consider an incoming brane. Then a straightforward change of
variables allows equation (\dr{parametric}) to be written in the
form $\phi = \int_{\wp}^{\infty} dx (4 x^3 - g_2 x - g_3)^{-1/2}$
where ${\wp}(\phi,g_2,g_3)$ is the Weierstrass function \cite{GR}.
Thus
\begin{equation}
\frac{r^2}{L^2}= y^2 = \frac{\hat{\ell}^2}{2qE} \left[ \frac{1}{3}
+ {\wp }\left(\phi ,\; g_2, \; g_3\right) \right]
\dle{solution}
\end{equation}
where
\begin{equation}
g_2=- 4 q E \left(\frac{\tE^2}{\hat{\ell}^4}-\frac{1}{6qE}
\right),\qquad g_3=- 4qE \left(\frac{2r_0^4E^2}{\hat{\ell}^4} +
\frac{\tE^2 q E}{\hat{\ell}^4} - \frac{1  }{27} \right).
\end{equation}
The properties of the elliptic function $\wp(\theta,g2,g_3)$ depend
on the sign of the discriminant
\begin{equation}
\Delta = g_2^3 - 27 g_3^2.
\end{equation}
Since $g_2$ and $g_3$ are both $\hat{\ell}$ and $E$ dependent,
imposing $\Delta=0$ for a given $E$ will determine a given angular
momentum:  this is the critical angular momentum $\ell_c$
mentioned above and shown in figure \dr{fig1}.  For large $E\gg
r_0$ (or equivalently in flat AdS$_5\times S_5$) we find
\beq
\Delta \sim (\ell^4-8 E^3)
\eeq
so that $\ell_c \sim 8^{1/4} E^{3/4}$ --- the exact  expression
valid for all $r_0$ is given in the appendix of \cite{SP}. For $\hl
> \hl_c$, $\Delta > 0$ and
$\wp(\theta,g2,g_3)$ is a periodic function whose period
determines the scattering angle $\phi_s$ at which the outgoing
brane is scattered off the centrifugal barrier. For $\Delta < 0$
the brane may reach the horizon.  The parametric plot of $r(\phi)$
given in (\dr{solution}) is shown in figure \dr{fig2}a where the
blackhole horizon is the solid circular line. When $\ell < \ell_c$
the orbiting brane is seen to spiral into the blackhole horizon in
a finite $\tau$-time. The other curves are branes with $\ell >
\ell_c$ which follow hyperbolic trajectories with decreasing
eccentricity as $\ell$ increases.  In the critical case $\ell =
\ell_c$ the brane reaches the circular orbit whose radius $r_c$
can be determined from the properties of $\wp$ for $\Delta=0$:
\beq
\frac{r_c^2}{L^2} = \frac{\hat{\ell}_c^2}{2qE} \left[ \frac{1}{3} -
\frac{3}{2}\frac{g_3}{g_2} \right] \qquad {\rm where} \qquad g_2^3
= 27 g_3^2.
\eeq
Finally, one can use (\dr{Rdot}) and (\dr{solution}) to obtain
$\eta(\phi)$ where $\eta$ is brane conformal time:
\beq
\eta = \frac{1}{\hl} \int_0^{\phi} y^2(\phi) = \frac{\hl}{2qE}
\left[ \frac{\phi}{3} + \zeta(\phi; g_2,g_3) \right],
\eeq
where $\zeta$ is the Weierstrass zeta function \cite{GR}.  In
figure \dr{fig2}b we have converted back to $\tau$-time and
plotted the brane scale factor $y(\tau)$ for the different
trajectories of figure \dr{fig2}a.  The bounce is clearly observed
when $\ell>\ell_c$ .
\begin{figure}
\centerline{ \scalebox{0.8}{
\includegraphics{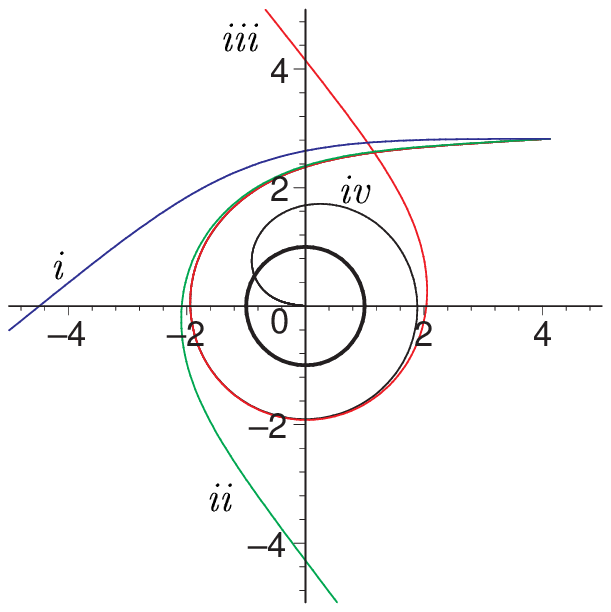}}
\hspace*{1cm} \raisebox{0.5cm}{
 \scalebox{0.8}{ \includegraphics{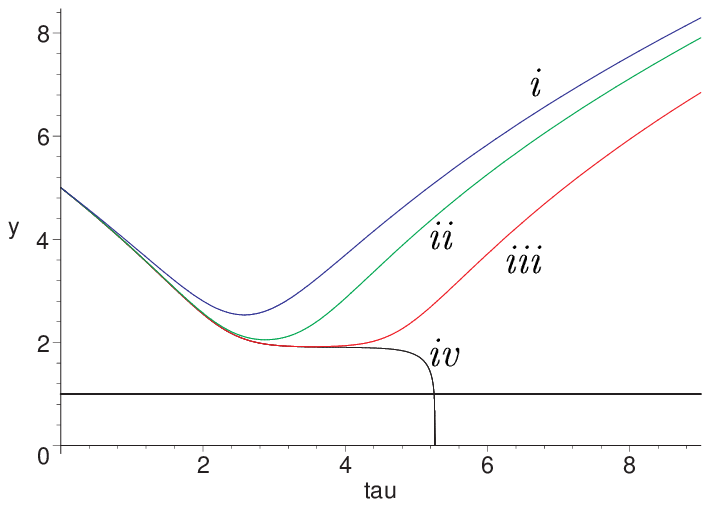}}}}
\caption{In both figures $q=+1$.  LH panel: The brane trajectory
in a flat universe $k=0$, corresponding to the potential of figure
1a. (Parametric plot of $y(\tau)$ and $\phi(\tau)$, with initial
condition placing the ingoing brane at $(4,3)$.). The central dark
circle is the blackhole horizon at $r_0=1$.  The trajectory of a
brane with a large angular momentum (blue curve, labeled {\it i})
is only slightly deflected by the blackhole.  As $\hl_c$ decreases
(green and then red curve, labeled {\it ii} and {\it iii}
respectively) the trajectory is more and more eccentric. For $\hl
< \hl_c$ the brane falls into the blackhole horizon. This is
represented by the black curve (labeled {\it iv}) which has $\ell
\lsim \ell_c$. RH panel: corresponding plot of the brane scale
factor $y(\tau)$ clearly showing the bounce.}
\dle{fig2}
\end{figure}
\\
\\
{\it Closed universe $k=1$.}
\\
As noted above and seen in figure \dr{fig1}b, it is now possible
to have a cyclic universe if $\ell> \ell_c$.  The corresponding
brane trajectory and scale factor for such a cyclic universe is
shown in figure \dr{fig3}, which has been determined numerically
since the equations of motion are no longer integrable. Generally
these cyclic trajectories do not close on themselves. A further
important difference between this closed case and the $k=0$ one
discussed above comes from the quartic nature of the denominator
of (\dr{parametric}). There may now be different pairs of repeated
roots
as can be seen from the upper curve of figure \dr{fig1}b which has
$\ell > \ell_m$:  in this regime $V^{\tau}_{\rm eff}>E$ for all
$y$ except in a very small vicinity of the horizon. Thus a bounce
can only occur for $\ell_c < \ell < \ell_m$.
\begin{figure}
\centerline{ \scalebox{0.8}{
\includegraphics{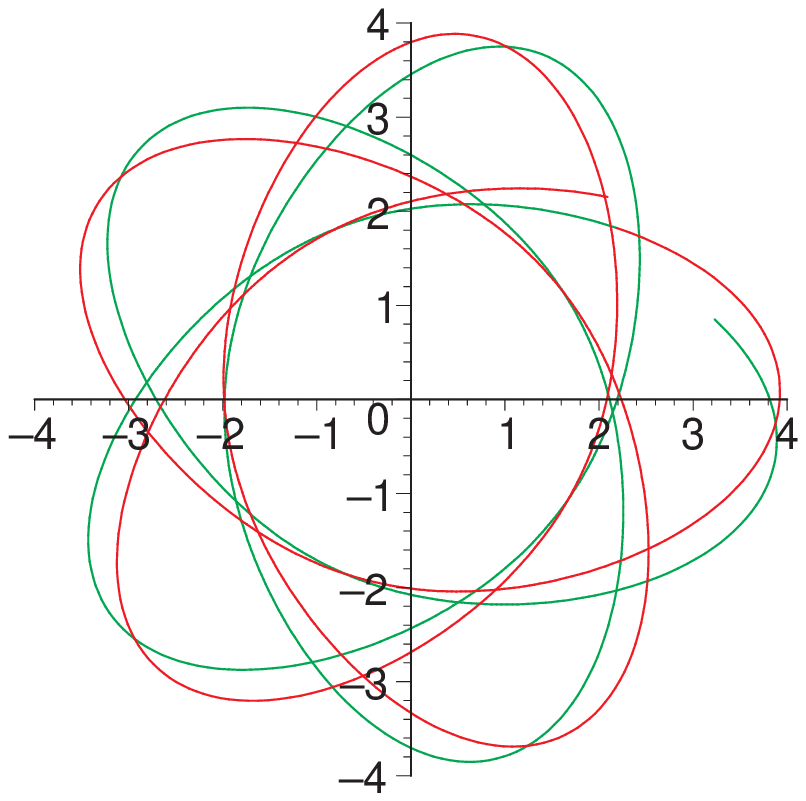}}
\hspace*{1cm} \raisebox{0.5cm}{
 \scalebox{0.8}{ \includegraphics{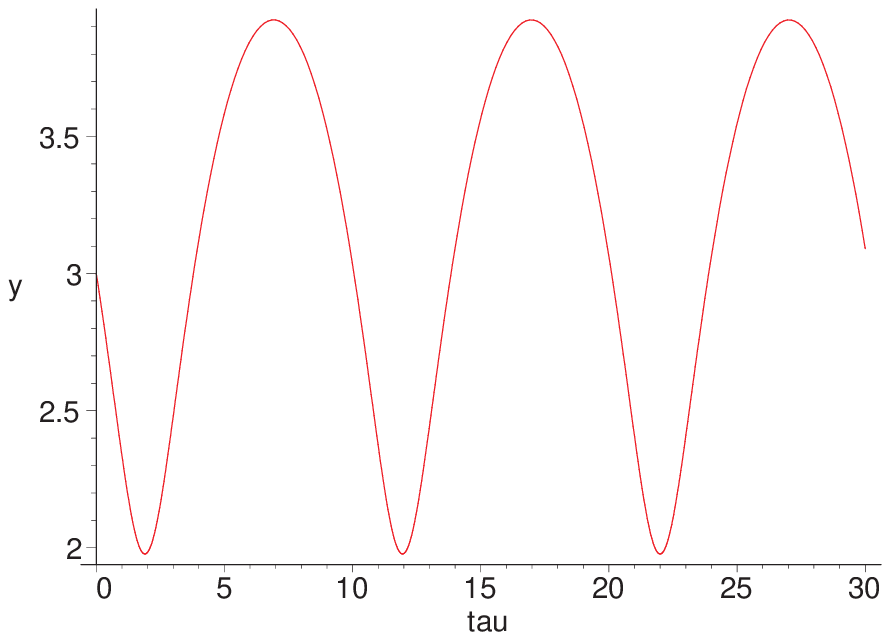}}}}
\caption{LH panel: The cyclic brane trajectory in a closed
universe $k=1$, corresponding to the potential of figure 1b.  Here
only one trajectory is drawn---in red, the trajectory from
$\tau_0=0$ to $\tau_0+50$, in green the trajectory from
$\tau_0+50$ to $\tau_0+100$.  The trajectories are generally not
closed.  RH panel: $y(\tau)$ for the brane.  The cyclic nature is
clearly observed. } \dle{fig3}
\end{figure}
\\
\\
{\it Comments on $AdS_5 \times S_5$.}
\\
An interesting limit to take is $r_0 =0$, namely a brane moving in
AdS$_5 \times S_5$.  Let us focus in particular on the static
limit $E=0$. Then the effective potential $V^{\tau}_{\bf eff}$ of
(\dr{Veffeta}) is given by
\beq
V^{\tau}_{\bf eff} = \frac{1}{2} \left[ k +
\frac{\hl^2}{y^6}(y^2+k) \right].
\dle{Vads}
\eeq
Thus, as expected, the potential is flat when $\hl =0$.  In that
case, the brane has no kinetic energy (i.e.\ is static) when
$k=0$, and constant kinetic energy when $k=-1$. In the closed case
$V^{\tau}_{\bf eff} > E$ always.

When $k=-1$ there are non-trivial effects when $\hl \neq 0$. From
(\dr{fdef}) the horizon is at $y_h=1$ where $V^{\tau}_{\bf
eff}=-1/2$. Once more the coefficient of $\hl^2$ is positive so
that we expect a centrifugal barrier.  Indeed the potential is
similar to that of figure \dr{fig1}a and again we find a bouncing
universe with $\hl_c = \sqrt{27}/2$.
\\
\\
{\it Asymptotically Flat Space-time}
\\
There is a large class of spacetimes for which bouncing branes
generically occur.  Consider the dynamics of a brane probe
embedded in an asymptotically flat spacetime whose isometry group
is assumed to contain $SO(6)$. Thus the background metric (in the
Einstein frame) is
\begin{equation}
ds^2=e^{2A(r)}(-dt^2+ d\vec{x}^2) +e^{2B(r)}(dr^2
+r^2d\Omega_5^2),
\end{equation}
where we also assume that the dilaton $\Phi(r)$ and the background
RR field $C_{0123}(r) \equiv e^{\Lambda(r)}$ depend only on $r$
and vanish at infinity. We take the coordinate system to be valid
with $r\in [r_H,\infty]$ where $r_H$ is an horizon such that
$e^{2A(r_H)}=0$. Then from (\dr{abcddef})
\begin{equation}
a = e^{8A} , \qquad b = - e^{2(3A + B)} = \frac{c}{r^2} , \qquad e
= - q e^{\Lambda}, \dle{abcddefbis}
\end{equation}
$\dot{r}^2$ is given in (\dr{Rdot}), and the brane scale factor
is $S = e^{\Phi/4}e^{A}$.

Rather that studying $V^{\tau}_{\rm eff}$ as above, let us
consider for simplicity $V^{t}_{\rm eff}(r,\ell,E) = E -
\dot{r}^2/2$ which contains the same information, though now as
seen by a bulk observer.
First of all notice that
$$V^{t}_{\rm eff}(r_H,\ell,E)=E $$
{\it i.e.\ }the kinetic energy as seen from an observer at
infinity vanishes at the horizon (the brane takes an infinite
$t$-time to fall into the horizon). When the probe is very far
from the horizon, $r \rightarrow \infty$, it is straightforward to
obtain that
$$ \dot{r}^2 \rightarrow 1-\frac{1}{E^2}. $$
Hence  the probe will be unable to escape to infinity if $E<1$.
This is very similar to the $k=1$ case studied above in the
explicit example of Sch-AdS$_5 \times S_5$.
On the other hand, if $E>1$ the probe will be able to escape. The
limiting case is when $E=1$, for which the kinetic energy of the
probe vanishes at infinity where there is no force on the probe.
Let us now consider a finite value of $r>r_H$. As $\ell$ goes to
infinity, the term in $V^{t}_{\rm eff}(r,\ell,E)$ proportional to
$\ell^2$ dominates implying that the potential becomes large and
positive over a finite interval in $r$ for a large enough value
$\ell> \ell_c$, just as in the example of Sch-AdS$_5 \times S_5$.
This implies that, for $E<1$ and $\ell>\ell_c$, there will be
cyclic solutions as in Fig.~\dr{fig2}.  On the other hand, for $E>1$ and $\ell>\ell_c$ the brane
motion possesses a branch of solutions where the brane is forced
to be between a finite value $r_{bounce}$ and infinity, cf.\
Fig.~\dr{fig1}. The value of $r_{bounce}$ is the largest positive
root of $V^{t}_{\rm eff}(r,\ell,E)=E$. It depends both on $\ell$
and $E$ but is guaranteed to exist as long as $E>1$ and
$\ell>\ell_c$. For a generic background the vanishing of $\dot
r^2$ at the bounce is linear $\dot r^2\sim (r-r_{bounce})$
implying that both in bulk time and proper brane time the bounce
takes place in a finite amount of time. Notice that the bounce is
intimately linked to the existence of a conserved quantity $\ell$
and therefore on the presence of more than one extra-dimension.

Finally let us notice that the bouncing branes that we have considered
are not singular. Hence they  should be amenable to a full treatment
 of the cosmological perturbations \cite{dani} before and after the bounce
\cite{jer,brand,gor}.
\\
\\
{\it Acknowledgements}  We thank C.~Charmousis for useful
discussions and T.~Evans who also helped a lot with Maple.

\typeout{--- No new page for bibliography ---}




\begin{thebibliography}{100}
\bibitem{Neil1} J.~Khoury, B.A.~Ovrut, P.J.~Steinhardt
and N.~Turok {\em The Ekpyrotic Universe: Colliding Branes and the
Origin of the Hot Big Bang}, Phys.\ Rev.\ {\bf D64} (2001) 123522.

\bibitem{Neil2} J.~Khoury, B.A.~Ovrut, N.~Seiberg, P.J.~Steinhardt
and N.~Turok {\em From Big Crunch to Big Bang}, Phys.\ Rev.\ {\bf
D65} (2002) 086007.

\bibitem{ovr} A.~Lukas, B.A.~Ovrut, K.S.~Stelle, D.~Waldram,
{\em Heterotic M-theory in Five Dimensions}, Nucl.\ Phys.\ {\bf
B552} (1999) 246.


\bibitem{Peloso} S.~ Mukherji and M.~Peloso, {\em Bouncing and cyclic
universes from brane models}, {\tt hep-th/0205180}.

\bibitem{Kackru}  S.~Kackru and L.~McAllister, {\it Bouncing brane
cosmologies from warped string compactifications}, {\tt
hep-th/0205209}.

\bibitem{KK}
A.~Kehagias and E.~Kiritsis, {\it Mirage cosmology}, JHEP {\bf 11}
(1999) 022,{\tt hep-th/9910174}.


\bibitem{Neil3}  P.J.~Steinhardt and N.~Turok, {\it A Cyclic Model of
the Universe}, {\tt hep-th/0111030};  {\it Cosmic Evolution in a
Cyclic Universe}, Phys.\ Rev.\ {\bf D65}  (2002) 126003.



\bibitem{SP}
D.~A. Steer and M.~F. Parry, {\em Brane cosmology, varying speed
of light and
  inflation in models with one or more extra dimensions},
{\tt hep-th/0201121}.


\bibitem{CU}
B.~Carter and J.-P. Uzan, {\em Reflection symmetry breaking
scenarios with
  minimal gauge form coupling in brane world cosmology}, Nucl. Phys. {\bf B606}
  (2001) 45, {\tt gr-qc/0101010}.


\bibitem{GR} {\it e.g.} I.S.Gradshtyn and I.M.Ryzhik, {\it Table
of integrals, series, and products}, Academic Press (5th edition),
London, 1984, pp 922.




\bibitem{dani} T.~Boehm and  D.A.~Steer, {\it Perturbations on a
moving D3-brane and mirage cosmology}, {\tt hep-th/0206147}.

\bibitem{jer} J.~Martin, P.~Peter, N.~Pinto Neto, D.J.~Schwarz,
{\it Passing through the bounce in the ekpyrotic models}, Phys.\
Rev.\ {\bf D65} (2002) 123513.


\bibitem{brand}
S.~Tsujikawa, R.~Brandenberger and  F.~Finelli, {\it On the
Construction of Nonsingular Pre-Big-Bang and Ekpyrotic Cosmologies
and the Resulting Density Perturbations}, {\tt hep-th/0207228}.

\bibitem{gor} C.~Gordon and  N.~Turok, {\it
Cosmological Perturbations Through a General Relativistic Bounce},
{\tt hep-th/0206138}.


\end{thebibliography}
\end{document}